\documentclass[]{nature_cw}
\usepackage{amssymb}
\usepackage{graphicx}
\usepackage{float}
\renewcommand{\thefigure}{\textbf{\arabic{figure}}}
\renewcommand{\figurename}{\textbf{Fig.}}

\title{A Room-temperature Ultrafast Spin-polarized Polariton Laser}
\author{Feng-kuo Hsu$^1$, Yi-Shan Lee$^2$, Sheng-Di Lin$^2$, and Chih-Wei Lai$^1$}

\begin{document}

\maketitle
\begin{affiliations}
\item Department of Physics and Astronomy, Michigan State University, East Lansing, MI 48824, USA
\item Department of Electronics Engineering, National Chiao Tung University, Hsinchu 30010, Taiwan 
\end{affiliations}

\begin{abstract}
Light-matter interactions are greatly altered in resonators and heterostructures where both electrons and photons are confined. These interactions lead to nonlinear optical processes and lasing where both photon mode control and electron quantum confinement play a role. Lasing in semiconductors is generally independent of the spins of electrons and holes, which constitute the gain medium. However, in a few spin-controlled lasers\cite{rudolph2003, blansett2005, holub2007, iba2011, gerhardt2012}, spin-polarized carriers with long spin relaxation times ($\gtrsim$ 1 ns) result in continuous or sub-nanosecond pulsed circularly polarized stimulated emission. In principle, optical or electrical injection of spin-polarized carriers can reduce the lasing threshold\cite{oestreich2005} and switching time\cite{zutic2011}. Here, we demonstrate room-temperature spin-polarized ultrafast (sub-10-ps) pulsed lasing in a highly optically excited GaAs microcavity embedded with InGaAs multiple quantum wells (MQWs) within which the spin relaxation time is less than 10 ps. The polariton laser produces fully circularly polarized radiation. By suppressing the steady-state thermal heating and tuning the MQW bandgap to the cavity resonance, we obtain a lasing efficiency greater than 10$\%$, matching the fraction of carriers photoexcited in the MQWs. Our results provide insights into Coulomb many-body effects in light-matter hybrids\cite{comte1982,nozieres1982,nozieres1985,fischer2013, kamide2011}.
\end{abstract}

The absorption of circularly polarized light in semiconductors results in the spin polarization of electrons (holes) in the conduction (valence) band, as determined by the optical selection rules\cite{wu2010}. The spin polarization decays on the scale of the spin relaxation time, $\tau_s$. Spin-polarized carriers with a long $\tau_s$ have been considered necessary for obtaining a spin-controlled laser (spin-laser). In cubic III-V semiconductors such as GaAs and InAs, holes typically have $\tau_s$ on the order of sub-picoseconds because of the strong spin-orbit interaction and valence-band mixing\cite{hilton2002}. Spin-polarized conduction band electrons have thus been the focus of potential spin-controlled optoelectronic devices and lasers. Recently, room-temperature circularly polarized lasing with 0.2 ns duration has been demonstrated in a vertical-cavity surface-emitting laser (VCSEL) with (110)-oriented GaAs/AlGaAs MQWs\cite{iba2011}, in which $\tau_s >$ 0.7 ns due to substantial suppression of a spin relaxation mechanism [D'yakonov-Perel' (DP) mechanism] predominant in conventional (100) QWs\cite{ohno1999}. A $\tau_s$ on the order of nanoseconds results in partially circularly polarized luminescence even below the lasing threshold. However, the growth  of high quality quantum well and microcavity structures on an unconventional (110) GaAs substrate remains challenging. In contrast, here, we describe room-temperature spin-polarized sub-10-ps pulsed lasing despite a much shorter electron $\tau_s \lesssim$ 10 ps in the InGaAs/GaAs MQWs\cite{tackeuchi1999} of our microcavity grown on a (100) GaAs substrate. Spin-polarized carriers are optically injected by non-resonant picosecond pump pulses. Below threshold, luminescence is unpolarized because $\tau_s<10$ ps is short compared to the carrier lifetime $\tau_n>1$ ns. At a critical photoexcited density, polariton laser action commences with a high degree of circular polarization, close to unity  ($>$0.98). We refer to such a spin-polarized laser as a polariton laser because it involves the mixing of the electronic polarization with the cavity light field at high photoexcited densities. Unlike a conventional semiconductor laser or VCSEL\cite{iga2008}, our polariton laser exhibits a nonlinear input-output relation, energy shifts, and spectral broadening as a function of the optical pump flux (number of photons per pulse). Analysis of polarization and spectral characteristics may lead to deeper understanding of Coulomb many-body effects\cite{schmitt-rink1985,schmitt-rink1985a, jahnke1995} and coupling of the electronic polarization to the cavity light field. 

The $\lambda$ GaAs/AlAs distributed Bragg reflector (DBR) microcavity studied here has three stacks of three InGaAs/GaAs MQWs each, embedded at the antinodes of the cavity light field (Fig. 1a and Methods). Five samples with energy detuning ($\Delta = E_X - E_C$) ranging from about $-5$ to $+5$ meV are investigated (Supplementary Fig. S3 and S7), where $E_X$ is the QW bandgap and $E_C$ is the cavity resonance. We do not consider exciton-polaritons in the strong coupling regime\cite{houdre2002, gibbs2011}, even though room-temperature exciton-polariton formation has been reported in a similar InGaAs-MQW-based planar microcavity\cite{houdre1994}. Excitons (bound electron-hole pairs) in the InGaAs MQW and GaAs spacer layers have a binding energy of about 8 meV and 4.2 meV, respectively, relatively small compared to the 25-meV room-temperature thermal energy. At room temperature, the photoexcited electron-hole system in our InGaAs MQWs consists mainly of free carriers or plasma owing to thermal ionization\cite{schmitt-rink1985a}. However, free-carrier polarization can also be coupled to a strong cavity light field. When laser action commences, the cavity light field increases dramatically. The coupling of free-carrier polarization to such an intense cavity light field can affect the lasing dynamics and characteristics. To investigate such polariton laser action in a highly optically pumped microcavity, we vary the laser pump flux by two orders of magnitude and create a photoexcited density ranging from approximately $5\times10^{11}$ to $10^{13}$ cm$^{-2}$ per QW per pulse. At such a high pump flux, the steady-state incident power transmitted to the sample at the 76 MHz repetition rate of the laser would exceed $50$ mW, resulting in significant thermal heating (Supplementary Fig. S4) and carrier diffusion (Supplementary Fig. S5). To control both heating and diffusion, we use temporal intensity modulation and spatial beam shaping techniques (Methods). 

We optically pump the sample using 2-ps Ti:Sapphire laser pulses at 1.579 eV ($\lambda_p = 785$ nm), which is near a reflectivity minimum (reflectance $\approx$40$\%$) of the microcavity and 170 meV above the cavity resonance (Fig. 1b). In InGaAs MQWs, the electron and hole spin relaxation times are short ($\tau_s\lesssim 10$ ps and $<1$ ps\cite{damen1991, tackeuchi1999}) compared to the carrier lifetime ($\tau_{n} >1$ ns). Optically injected spin-polarized carriers in InGaAs MQWs are expected to lose spin polarization after dissipating 100 meV of excess energy. Indeed, luminescence under a circularly polarized pump below the threshold is unpolarized. Surprisingly, circularly polarized ultrafast pulsed laser action commences above the threshold photoexcited density $n_{th}\approx3 \times 10^{12}$ cm$^{-2}$ per QW. The lasing commences within 10 ps at an energy between those of the QW bandgap and cavity resonance. The 2D $n_{th} $ in a 6-nm thick QW corresponds to an effective 3D density $N_{th} = 5 \times 10^{18}$ cm$^{-3}$ or an effective threshold current density $j^*_{th}\approx$ 500~A~cm$^{-2}$, assuming $j^*_{th} \approx {n_{th}\, e}/ {\tau_{n}}$ and carrier lifetime $\tau_{n} = 1$ ns. The effective threshold density is comparable to that of an electrically pumped VCSEL\cite{iga2008} or polariton laser\cite{schneider2013}.

We now wish to distinguish the present room-temperature system from exciton-polariton condensates at cryogenic temperatures\cite{kasprzak2006, lai2007, deng2010, deveaud-pledran2012, carusotto2013}. A laser is typically characterized by its coherence and spectral properties, while a macroscopic condensate exhibits unique energy and momentum distributions. Fig. 2 contains data showing the increase of spatial coherence, nonlinear growth of macroscopic occupation in a state in energy and momentum space, and the spontaneous increase of circular polarization. We note that some exciton-polariton condensates exhibit linearly polarized radiation owing to an energy splitting of the order of $100$ $\mu$eV between two linearly polarized modes ($\sigma^X$ and $\sigma^Y$) induced by structural disorder and strain\cite{kasprzak2006}. The angular and spectral distributions of radiation from the polariton laser studied here resemble those observed in the condensates of exciton-polaritons at cryogenic temperatures\cite{kasprzak2006, lai2007}. However, we refer to our room-temperature microcavity system as a laser because of its dynamic nature and the absence of exciton-polaritons in the strong coupling regime at high photoexcited densities\cite{houdre2002}. First, we characterize our polariton laser in terms of the angular distribution and energy as functions of the in-plane momentum [angle-resolved (k-space) images and $E$ vs. $k_\parallel$ dispersions] (Methods). In a planar microcavity, carriers coupled to the cavity light field are characterized by an in-plane wavenumber $k_\parallel = k \, \sin (\theta)$ owing to the two-dimensional confinement of both photons and carriers. The leakage photons can thus be used for a direct measurement of the angular distribution of optically active carriers. A detailed analysis of k-space images and spatial coherence is provided in Supplementary Fig. S6. In Fig. 2a, we show the selected k-space luminescence images and spectra of Sample-L1406, which has a lasing energy of 1.406 eV at threshold. Far below threshold, luminescence from GaAs layers dominates and forms an isotropic angular distribution (not shown). Slightly below threshold, a parabolic $E$ vs. $k_\parallel$ dispersive spectrum appears near the bandgap energy of the InGaAs MQWs, characteristic of the mixing of the free-carrier polarization and cavity light field (Supplementary Fig. S3). At threshold, the radiation becomes angularly and spectrally narrow. An intense radiation mode emerges within an angular spread $\Delta\theta<3^\circ$, corresponding to a standard deviation in k-space $\Delta k = 0.3$ $\mu$m$^{-1}$. Approximating such a partially coherent beam as a Gaussian Schell-model source\cite{friberg1982}, we can determine a spatial coherence length of 4 $\mu$m, close to the dimension of the spatial mode. Above threshold, the energy of radiation blueshifts with increasing spectral linewidth while the radiation remains highly directional with the pump flux. Next, we describe the nonlinear input-output relation. Fig. 2b shows emission flux (output) vs. pump flux (input) under circularly ($\sigma^+$) or linearly ($\sigma^X$) polarized excitation. Pump flux, $P$, is the photon flux per pulse transmitted into the microcavity within a circular 10-$\mu$m diameter area. The output nonlinearly increases by one order of magnitude for an increase of the input by $<20\%$ near the critical photoexcited density. The onset of such a nonlinear output for the co-circular component ($\sigma^+/\sigma^+$) is defined as the threshold $P_{th}$ (indicated by an arrow in the figure). The threshold pump flux can vary up to $50\%$ owing to spatial inhomogeneities and disorder (Supplementary Fig. S7). For $P_{th}\approx 2.5 \times 10^8$ per pulse over an area of 80 $\mu$m$^2$, $n_{th}$ for each QW is estimated to be $3 \times 10^{12}$ cm$^{-2}$. For a given photoexcited density $n_{th}=3 \times 10^{12}$ cm$^{-2}$, the dimensionless mean separation of carriers\cite{kamide2011} $r_s \equiv 1/(\sqrt{n_{th}} \, a_B) \approx 0.6$, where  $a_B \approx 10$ nm is the exciton Bohr radius in an In$_{0.15}$Ga$_{0.85}$As QW. For $P>1.2\,P_{th}$, the cross-circularly polarized component also lases. Under a linearly polarized pump, the laser action commences at a slightly higher pump flux ($P = 1.05 \, P_{th}$). This 5$\%$ threshold reduction with optical injection of spin-polarized carriers is small but significant compared to the $<$1$\%$ reduction predicted for an InGaAs-MQW-based VCSEL\cite{oestreich2005}. In general, such a threshold reduction is less than 5$\%$ in most locations and samples studied here. We define the efficiency as the ratio of the emission flux emanating from the front surface (output) to the pump flux transmitted into the microcavity (input). In this sample, the total emission under a circularly polarized pump is close to that under a linearly polarized one. The overall efficiency reaches a plateau of $3.5\%$ at $P\gtrsim3\,P_{th}$. In the plateau regime, the output linearly increases with the input, resembling the characteristics of a conventional semiconductor laser. Maximal efficiency ranges from $3\%$ to $11\%$. An efficiency greater than $10\%$ is obtained in Sample-L1410, in which $\Delta \approx 0$ at low photoexcited densities (Supplementary Fig. S8). Because optical absorption in the nine 6-nm thick In$_{0.15}$Ga$_{0.85}$As/GaAs MQWs in the cavity is $12\%$ at $\lambda_p=785$ nm at room temperature, an efficiency greater than $10\%$ implies that essentially all of the carriers photoexcited in the MQWs can recombine radiatively and contribute to laser action.

The spontaneous build-up of the circularly polarized radiation at a critical photoexcited density (Fig. 2c) can be quantified by the Stokes vector $S = \{S_0,S_1,S_2, S_3\}$ (Methods). We use three quantities deduced from the normalized Stokes three-vector $s = \{s_1, s_2, s_3\}$ to represent the polarization state: the degree of circular polarization ($DoCP=s_3$), degree of linear polarization ($DoLP=\sqrt{s_1^2+s_2^2}$), and degree of polarization ($DoP=\sqrt{s_1^2+s_2^2+s_3^2}$). Below threshold, the radiation is unpolarized as indicated by the zero value of $DoP$. At threshold, radiation with $DoCP$ greater than 0.95 spontaneously increases near $k_\parallel = 0$. In this slightly above threshold regime ($P_{th}\,<\,P\,<1.2\,P_{th}$), the radiation is highly circularly polarized with a slight ellipticity. With increasing pump flux, $DoCP$ gradually decreases to 0.25 at $P = 4\,P_{th}$, while $DoLP$ rapidly decreases to below $0.1$ for $P>1.5\,P_{th}$. Here, the radiation becomes elliptically polarized with reduced $DoP$. When the helicity of the circularly polarized pump is switched, $DoCP$ changes sign but maintains the same magnitude, i.e., the polarization state is symmetric with respect to the helicity of the pump. Under a linearly polarized pump, radiation with finite $DoLP$ and $DoCP$ between 0.5 to 0.2 (i.e., elliptical polarization) is observed only slightly above threshold for $P_{th}<P<1.2\,P_{th}$ and becomes unpolarized for $P> 1.5\,P_{th}$ (not shown). Depending on the locations of the sample, the polarization ellipse can be oriented towards the $[110]$ or $[1\bar{1}0]$ crystallographic direction of the GaAs layers. In contrast to parametric amplification processes, the polarization state of the present polariton laser is affected by but not locked to the pump polarization state. The spontaneous build-up of full circularly and partially linearly polarized radiation within only a limited density regime is indicative of a coherent process.

To understand the mechanism of the spin-polarized laser action, it is necessary to investigate the polarization dynamics through time-resolved polarimetry and spectroscopy. Fig. 3a shows selected time-resolved co- and cross-circularly polarized luminescence [$I^\pm(t)$] under $\sigma^+$ circularly polarized pump. Below threshold, the luminescence exhibits a long decay time $\tau_r>2$ ns and is unpolarized. At threshold, the co-circular component commences pulsed laser action within 50 ps, while the cross-circular component remains negligible [$I^{+}(t)/I^{-}(t)>100$] with a decay time $\tau_r>100$ ps. Above threshold, pulsed radiation becomes as short as 5 ps with a sub-10-ps rise time with increasing pump flux. For $P \gtrsim 1.5 P_{th}$, the cross-circular component also begins lasing. The rise and decay times of the co- and cross-circular components are shown in Supplementary Fig. S9. Under a non-resonant excitation above the bandgaps of the InGaAs QW (active layers) and GaAs (cavity spacer layers), electron-hole pairs can be photoexcited in both InGaAs MQWs and GaAs layers. If diffusion of carriers photoexcited in the GaAs cavity spacer layers into the InGaAS MQWs occurs, it is expected to take $\sim$100 ps in the absence of a cross-well electric field. Therefore, the sub-10-ps pulsed lasing suggests that radiation is dictated by carriers initially photoexcited in InGaAs MQWs. Next we consider the temporal $DoCP(t)$. Below threshold, $DoCP(t)$ is minimal even within the initial 10 ps, implying a sub-10-ps spin relaxation time and $\tau_s  \ll  \tau_r$. At threshold, $DoCP(t)$ reaches 0.98 and remains above 0.8 for more than 60 ps, indicative of an increase of the spin relaxation time at high photoexcited densities. Such a high $DoCP(t)$ through the pulsed duration suggests an ultrafast energy relaxation or stimulated process in which spin polarization is preserved. The coexistence of cross- and co-circular lasing makes it difficult to discriminate spin flipping from energy relaxation through these spectrally integrated time-resolved measurements. To differentiate spin flipping from energy relaxation, we conduct temporally and spectrally resolved polarimetric measurements, as shown in the streak images in Fig. 3b. Selected cross-sectional transient spectra are provided in Supplementary Fig. S10. At threshold, the radiation remains circularly polarized and spectrally narrow with a peak energy that is nearly constant with time. Above threshold, the radiation expands spectrally when the laser action commences and gradually redshifts with time. In semiconductor lasers, amplitude and phase fluctuations are known to modulate the index of refraction of the gain medium, leading to an increased linewidth\cite{henry1982,henry1986}. The spectral broadening with pump flux here is largely due to such fluctuations induced by carrier-carrier interactions. If it were due to a transient cavity resonance shift with photoexcited density, the transient spectra would remain narrow but redshift with time when the carriers decay from the cavity. 

In contrast to a conventional semiconductor laser, our polariton laser exhibits nonlinear energy shifts, spin-dependent energy splittings, and linewidth broadening with increasing pump flux, all due to Coulomb many-body interactions. Fig. 4a shows the co- and cross-circularly polarized spectra at $k_\parallel=0$ under a circularly polarized ($\sigma^{+}$) pump. The corresponding linewidth $\Delta E$ (standard deviation of the spectral distribution) and mean peak energy (center of mass of the spectral distribution) are plotted in Fig. 4b, while $DoCP(E)$ curves for the selected $P$ are shown in Fig. 4c. Far below threshold (regime I: $P<0.8\, P_{th}$), we see only spectrally broad luminescence ($\Delta E>30$ meV) from the GaAs spacer layers. Slightly below threshold (regime II: $0.8\,P_{th}< P< P_{th}$), luminescence from the radiative recombination of carriers in InGaAs MQWs becomes visible and gradually dominates over that from GaAs layers. In particular, a parabolic $E$ vs. $k_\parallel$ dispersive curve consisting of spectrally narrower peaks ($\Delta E\lesssim 5$ meV) appears near the bandgap of InGaAs MQWs (Fig. 2a and Supplementary Fig. S2). From luminescence spectra at $P=0.9\,P_{th}$ (not shown), we anticipate the linear anisotropic splitting to be below 100 $\mu$eV, limited by the spectral resolution. On the other hand, we observe a circular anisotropic splitting $\approx$400 $\mu$eV (Fig. 4b). When the pump flux is increased toward the threshold, luminescence blueshifts by $\approx$4 meV, while the linewidth decreases from about 5 meV to 0.3 meV. Slightly above threshold (regime III: $P_{th} < P\lesssim 1.5\,P_{th}$), spectrally narrow radiation emerges ($\Delta E\approx$0.3-1.0 meV) with a nonlinear growth in magnitude. In a limited density regime, the mean peak energy also remains constant with increasing pump flux. Far above threshold (regime IV: $P\gtrsim 1.5\,P_{th}$), the spectral linewidth increases to 2 meV because of the existence of multiple spectral components. Co- and cross-circular components both lase with rising mean peak energy while retaining an energy splitting of $\approx$1 meV. The energy shift up to 3 meV is due to mean-field interactions (many-body effects) of spin-polarized carriers. Such spin-dependent energy splittings of $\approx$1 meV are observed for all five samples with varying detuning $\Delta$, although the magnitudes vary slightly. Furthermore, the $DoCP(E)$ curves are highly dependent on the photoexcited density, suggesting that Coulomb many-body effects affect both the spin and energy relaxation processes. 

In semiconductor lasers, lasing modes are typically linearly polarized along the $[110]$ or $[1\bar{1}0]$ crystallographic directions due to anisotropic strain or disorder. But unwanted polarization switching between these modes can occur with increasing carrier densities, thermal heating, or in-plane anisotropic strain\cite{panajotov2013}. In contrast, the present highly optically excited microcavity can produce stable fully circularly polarized radiation. This stability may result from a spin-dependent energy splitting of about 1 meV (even in the absence of a magnetic field). The polarization properties of radiation are determined by the electron and hole spin relaxation processes instead of crystal anisotropy. Our polariton laser also shows a nonlinear input-output relation and energy shifts owing to the mixing of the free-carrier polarization and cavity light field. The microscopic mechanism of the spontaneous increase of spin polarization is probably related to the exclusion principle (phase-space filling/Pauli blocking)\cite{schmitt-rink1985a}. However, a microscopic model based on Maxwell-Bloch equations beyond a phenomenological spin-flip model\cite{san-miguel1995} is necessary to substantiate that hypothesis. Our results should stimulate further work into exploiting spin-orbit and many-body interactions for practical spin-dependent optoelectronic devices. 

\section*{Methods Summary}
The microcavity studied here consists of three stacks of three In$_{0.15}$Ga$_{0.85}$As/GaAs (6-nm/12-nm) multiple quantum wells each, embedded at the antinodes of the light field in a one-wavelength ($\lambda$) GaAs cavity (Supplementary Fig. S1). The cavity quality factor Q $\approx$ 4000-7000, corresponding to a cavity photon lifetime $\tau_c \approx 2$ ps. The techniques we use to control the heating and diffusion of photoexcited carriers are (1) temporally modulating the pump laser intensity to suppress steady-state thermal heating, and (2) spatially shaping the pump beam profile to enable lasing in a single transverse mode. To suppress stead-state thermal heating, we temporally modulate the 2-ps pump laser pulse train at 76 MHz with an on/off ratio (duty cycle) $<$1$\%$ using a double-pass acousto-optic modulator (AOM) system at 1 kHz. We limit the time-averaged power to below 0.2 mW for all experiments. To control carrier diffusion, we holographically generate a flat-top beam profile (area $\approx$ 300 $\mu$m$^2$) at the sample surface using a two-dimensional (2D) liquid-crystal spatial light modulator (SLM). We use a Fourier transform optical system for angle-resolved (k-space) luminescence imaging, spectroscopy, and polarimetry (Supplementary Fig. S2). The polarizations of excitation and emissions are controlled or analyzed using liquid crystal devices without mechanical moving parts. A circularly polarized pump or luminescence with angular momentum $+\hbar$ ($-\hbar$) along the pump laser wavevector $\hat{k} \parallel \hat{z}$ is defined as $\sigma^{+}$ ($\sigma^{-}$). 

\section*{Methods}
\textbf{Sample fabrication.} We grow the microcavity on a semi-insulating (100)-GaAs  substrate using a molecular beam epitaxy (MBE) method. A schematic of the structure and a cross-sectional SEM image of the active region are shown in Supplementary Fig. S1. The top (bottom) DBR consists of 17 (20) pairs of GaAs(61-nm)/AlAs (78-nm) $\lambda/4$ layers. The central cavity layer consists of three stacks of three In$_{0.15}$Ga$_{0.85}$As/GaAs (6-nm/12-nm) quantum wells each, positioned at the anti-nodes of the cavity light field. The structure is entirely undoped and contains a $\lambda$ GaAs cavity sandwiched by DBRs, resulting in a cavity resonance $\approx$ 1.41 eV ($\lambda_c$ = 880 nm) at room temperature (Supplementary Fig. S2). The QW  bandgap is tuned with respect to the cavity photon resonance through a rapid thermal annealing process (at 1010-1090$^\circ$C for 5-10 s), in which the InGaAs QW bandgap blueshits due to the diffusion of gallium ions into the MQW layers. In such an InGaAs-MQW-based microcavity, the optical absorption in DBRs is negligible near the lasing energy even at room temperature. The cavity quality factor Q $\approx E_L/ (\Delta E_m)\approx$ 4000-7000, where $\Delta E_m \approx 0.2$ meV is the minimal linewidth and $E_L = h \nu_L \approx 1.41$ eV is the radiation energy measured at threshold. The Q-factor corresponds to a cavity photon lifetime $\tau_c = Q / (2 \pi \nu_L) \approx 2$ ps.

\textbf{Thermal management.} The techniques we use to control the thermal heating and diffusion of photoexcited carriers are (1) temporally modulating the pump laser intensity to suppress the thermal carrier heating, and (2) spatially shaping the pump beam profile to enable lasing in a single transverse mode. Steady-state thermal heating can inhibit laser action and lead to spectrally broad redshifted luminescence (Supplementary Fig. S4). To suppress steady-state thermal heating, we temporally modulate the 2-ps 76 MHz pump laser pulse train with a duty cycle (on/off ratio) $<$ 0.5 $\%$ using a double-pass acousto-optic modulator (AOM) system. The time-averaged power is limited to below 0.2 mW for all experiments. Multiple transverse modes can simultaneously lase, owing to the diffusion of photoexcited carriers and crystalline disorder, leading to instability and complex lasing characteristics. To control carrier diffusion, we holographically generate a flat-top pump beam profile (area $\approx$ 300 $\mu$m$^2$) at the sample surface using a spatial beam shaper consisting of a two-dimensional (2D) liquid-crystal spatial light modulator (SLM) (Supplementary Fig. S5). 

\textbf{Experimental set-up: excitation.} 
The front surface of the microcavity is positioned at the focal plane of a high-numerical-aperture microscopy objective (N.A.=0.42, $50\times$, effective focal length 4 mm). A 3$\times$ telescope, a Faraday rotator, a polarizing beam splitter, and the objective form a reflected Fourier transform imaging system (Supplementary Fig. S2). The light fields at the SLM and sample surface form a Fourier transform pair. The 2D SLM ($1920 \times 1080$ pixels, pixel pitch = 8 $\mu$m) enables us to generate arbitrary pump geometries with a $\approx$2 $\mu$m spatial resolution at the sample surface using computer-generated phase patterns. The pump flux can be varied by more than two orders of magnitude using a liquid-crystal attenuator.
 
\textbf{Experimental set-up: imaging and spectroscopy.} 
We measure the angular, spectral, and temporal properties of  luminescence in the reflection geometry. Angle-resolved (k-space) luminescence images and spectra are measured through a Fourier transform optical system (Supplementary Fig. S2). A removable $f = 200$ mm lens enables the projection of either the k-space or r-space luminescence onto the entrance plane or slit of the spectrometer. Luminescence is collected through the objective, separated from the reflected specular and scattered pump laser light with a notch filter, and then directed to an imaging spectrometer. A single circular transverse lasing mode with a spatial mode diameter $\approx$8 $\mu$m is isolated for measurements through a pinhole positioned at the conjugate image plane of the microcavity sample surface. The spectral resolution is $\approx$0.1 nm (150 $\mu$eV), determined by the dispersion of the grating (1200 grooves/mm) and the entrance slit width (100-200 $\mu$m). The spatial (angular) resolution is $\approx$ 0.3 $\mu$m (6 mrad) per CCD pixel.

\textbf{Polarization control and notation.} The polarization state of the pump (luminescence) is controlled (analyzed) by a combination of liquid crystal devices, such as variable retarders and polarization rotators, and Glan-Taylor/Glan-Thomson polarizers without mechanical moving parts. A polarization compensator (Berek's variable wave plates) is used to compensate for the phase retardance induced by the reflection from the miniature gold mirror surface. The circularly polarized pump or luminescence with angular momentum $+\hbar$ ($-\hbar$) along the pump laser wavevector $\hat{k} \parallel \hat{z}$ is defined as $\sigma^{+}$ ($\sigma^{-}$). Linearly polarized light with horizontal (vertical) polarization is defined as $\sigma^X$ ($\sigma^Y$). The polarization state is characterized by the Stokes vector $\{S_0, S_1, S_2, S_3\}$. $S_0$ is the flux and is determined as $S_0 = I^{+}+I^{-} = I^{X}+I^{Y} = I^{45^\circ}+I^{135^\circ}$. The Stokes vector can be normalized by its flux $S_0$ to the Stokes three-vector $s = \{s_1, s_2, s_3\}$. $s_1 = (I^{X}-I^{Y})/(I^{X}+I^{Y})$, $s_2 = (I^{45^\circ}-I^{135^\circ})/(I^{45^\circ}+I^{135^\circ})$, and $s_3 = (I^{+}-I^{-})/(I^{+}+I^{-})$. $I^{+}$, $I^{-}$, $I^{X}$, $I^{Y}$, $I^{45^\circ}$, and $I^{135^\circ}$ are measured time-integrated or temporal intensities of the circular or linear polarized components. For unpolarized light $s = \{0, 0, 0\}$, $DoP = 0$; while for a completely polarized state, $DoP = 1$. The accuracy of the measured polarization state is $\approx$1-2$\%$.


\bibliographystyle{naturemag}
\bibliography{rtsppl}

\begin{addendum}
\item We thank Brage Golding, John A. McGuire, Carlo Piermarocchi, Y. Ron Shen, David Snoke and Hailin Wang for the discussions, and Jack Bass for critical reading of the manuscript. This work was supported by the NSF CAREER award DMR-0955944.
\item[Author Contributions] F.K.H. and C.W.L. performed the experiments and analyzed the data. Y.S.L. and S.D.L. grew the samples. C.W.L. conceived the project and wrote the manuscript with input from all authors. 
\item[Correspondence] Correspondence and requests for materials should be addressed to C.W.L. (cwlai@msu.edu).
\end{addendum}

\begin{figure}[H]
\centering
\includegraphics[width= 0.65 \columnwidth]{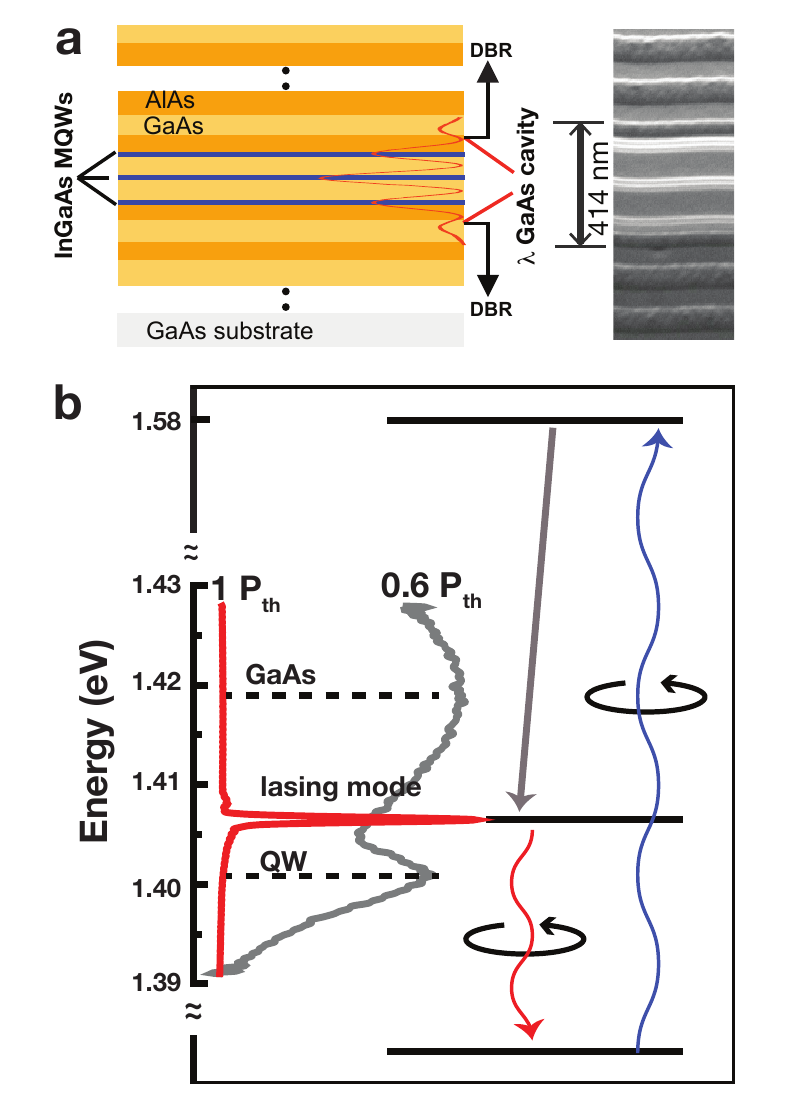}
\caption{\textbf{Schematics of the microcavity structure and optical excitation.} \textbf{a}, Schematic of the microcavity structures and a cross-sectional SEM image of the active region. \textbf{b}, Energy levels with normalized luminescence spectra below and at threshold.}
\end{figure}

\begin{figure}[H]
\centering
\includegraphics[width= 0.45 \columnwidth]{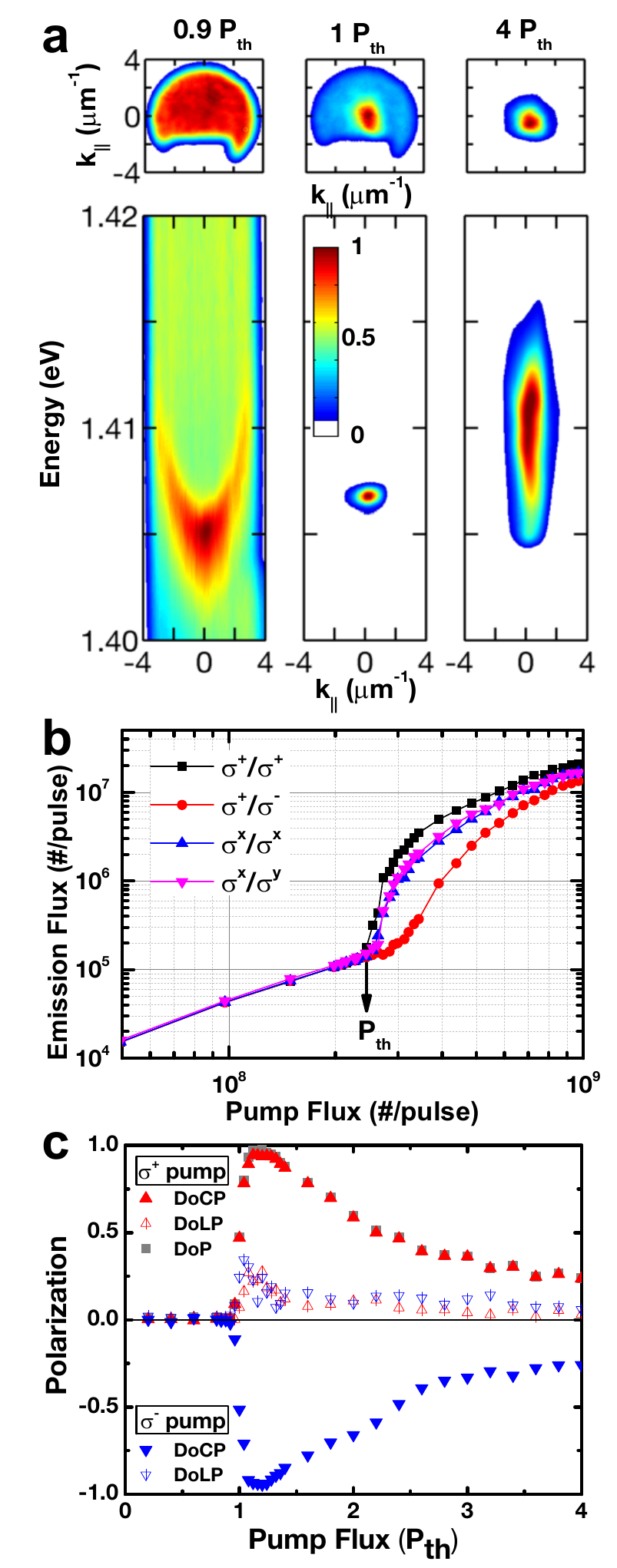}
\caption{\textbf{k-space imaging/spectroscopy and polarization states.} \textbf{a}, Angle-resolved (k-space) luminescence images and $E$ vs. $k_\parallel$ spectra at $P=$ $0.9$, $1$, and $4\,P_{th}$. The miniature mirror is positioned to direct the pump beam at an incident angle $\approx 20^\circ$ in front of the objective. The mirror partially blocks the reflected luminescence near a corner. \textbf{b},  Emission flux integrated over $|k_\parallel| \lesssim 3$ $\mu$m$^{-1}$ under $\sigma^+$ (circularly) or $\sigma^X$(linearly) polarized pump. \textbf{c}, The polarization state determined from luminescence integrated over $|k_\parallel|<0.3$ $\mu$m$^{-1}$ under $\sigma^+$ or $\sigma^-$ circularly polarized pump.} 
\end{figure}

\begin{figure}[H]
\centering
\includegraphics[width= 0.8 \columnwidth]{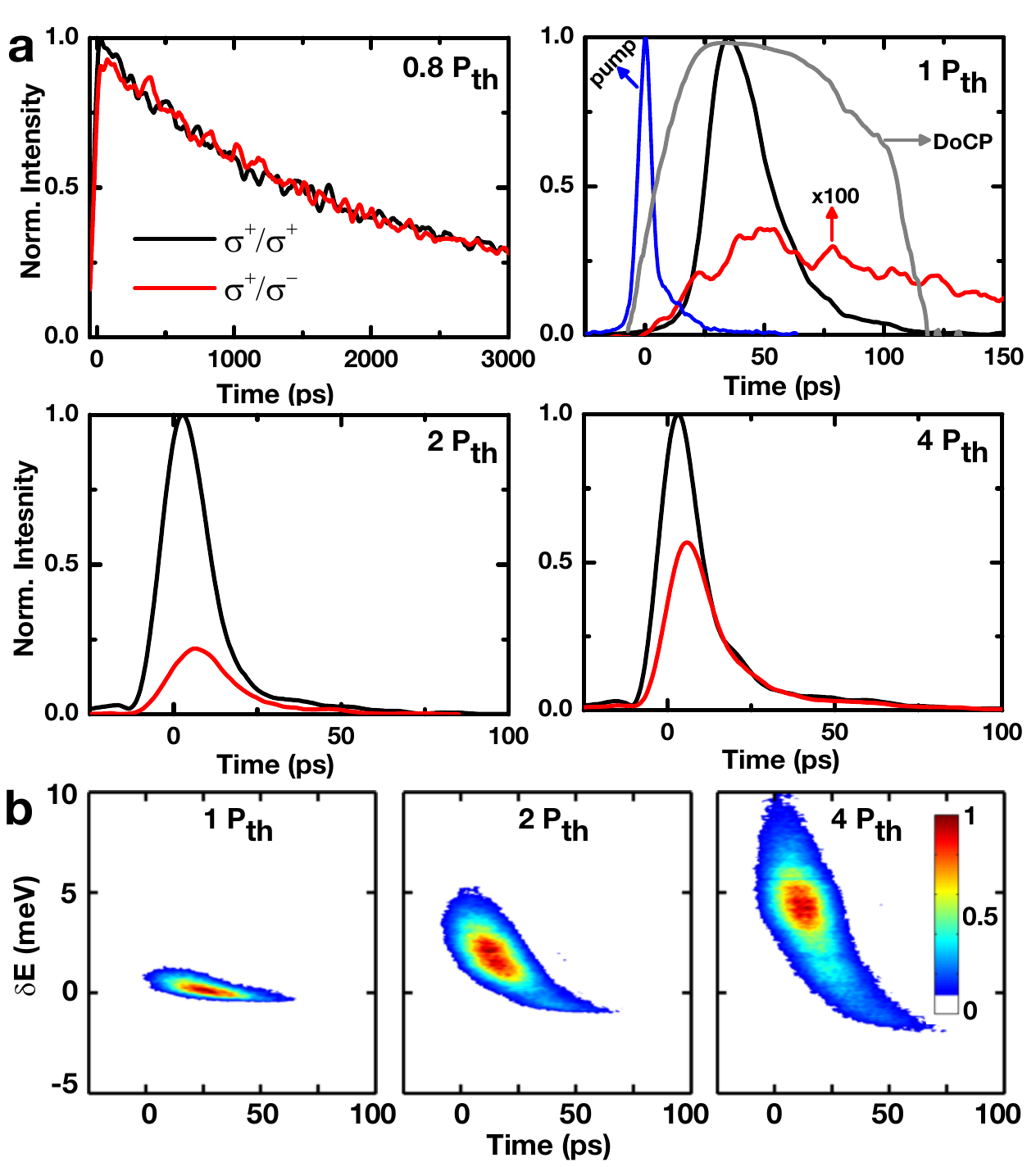}
\caption{\textbf{Dynamics and energy relaxation.} \textbf{a}, Time-resolved luminescence at $k_\parallel = 0$ for 0.8, 1, 2 and 4 $P_{th}$ under a circular polarized ($\sigma^+$) pump. Black (red) curves represent the co-circular $I^+(t)$ [cross-circular $I^-(t)$] components. Note that the cross-circular component shown at $P_{th}$ is multiplied by a factor of 100. The $DoCP(t)$ at $P_{th}$ is represented by the grey line. The time zero is determined from the instrument response (blue curve), which is measured via pump laser pulses reflected off the sample surface. The time traces are spectrally integrated, resulting in a temporal resolution $\approx 5$ ps.  \textbf{b}, Temporally and spectrally resolved streak images for 1, 2, and 4 $P_{th}$. The y-axis ($\delta E$) is offset with respect to $1.406$ eV ($882$ nm), the lasing energy at $P_{th}$. The temporal resolution is $\approx$30 ps owing to the grating-induced dispersion.}
\end{figure}

\begin{figure}[H]
\centering
\includegraphics[width= 0.6 \columnwidth]{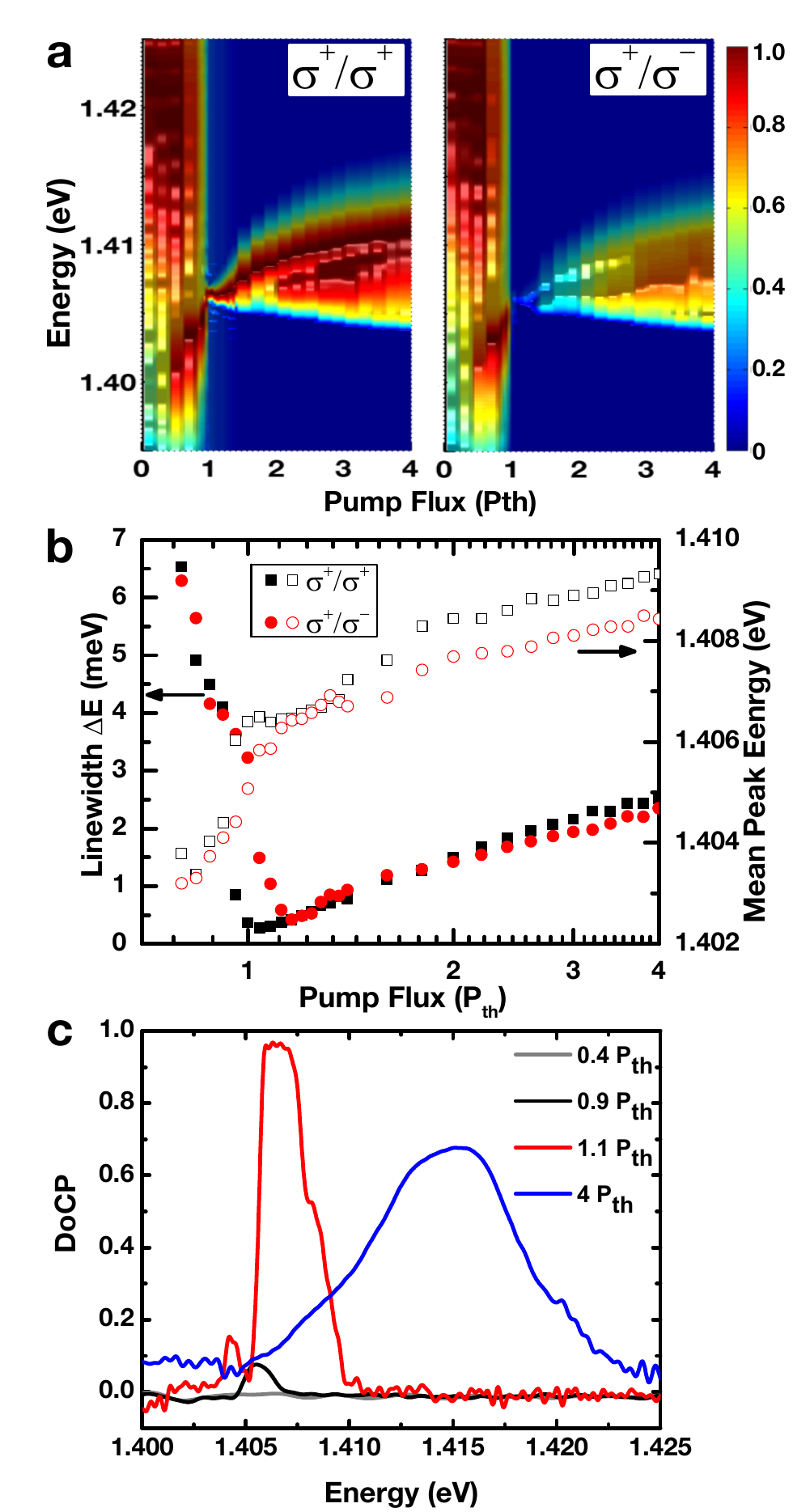}
\caption{\textbf{Spectral characteristics.} \textbf{a}, Spectra at $k_\parallel = 0$ vs. pump flux. The intensity $I(E)$ is normalized with respect to the co-circular $\sigma^+/\sigma^+$ component for each pump flux. The false color scale represents $\sqrt{I(E)}$. \textbf{b}, Spectral linewidths and mean energies determined from the spectra in \textbf{a}, with the luminescence from GaAs layers subtracted. \textbf{c}, $DoCP(E)$ at $P=$ 0.4, 0.9, 1.1, and 4 $P_{th}$.} 
\end{figure}

\newpage
\section*{\center\Large{Supplementary Information}}
Here, we provide additional experimental data and figures as Supplementary Information.
\setcounter{figure}{0} \renewcommand{\thefigure}{\textbf{S\arabic{figure}}}
\renewcommand{\figurename}{\textbf{Fig.}}
\begin{figure}[H]
\includegraphics[width= 1.0 \columnwidth]{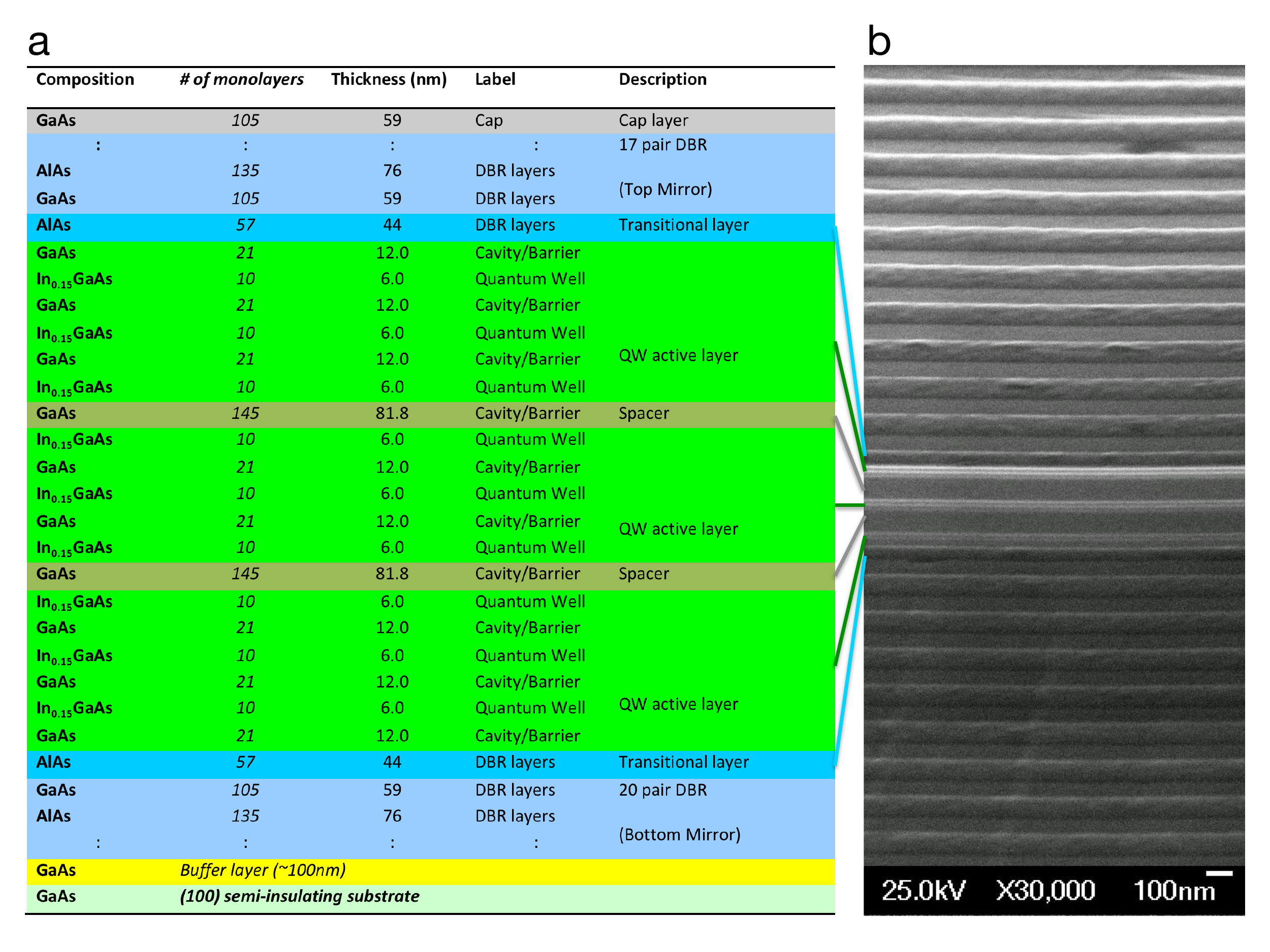}
\caption{\textbf{Microcavity structure.} \textbf{a}, Structure of the microcavity: InGaAs multiple quantum wells (MQWs) embedded within a $\lambda$ GaAS cavity. We adjust the antinodes of the cavity light field to the MQW layers by adding two transitional AlAs layers and a GaAs cap layer. \textbf{b}, A cross-sectional SEM image showing the active regime and adjacent layers of the distributed Bragg reflectors (DBRs).}
\label{fig:ucav_struc}
\end{figure}

\begin{figure}
\centering
\includegraphics[width= 0.8 \columnwidth]{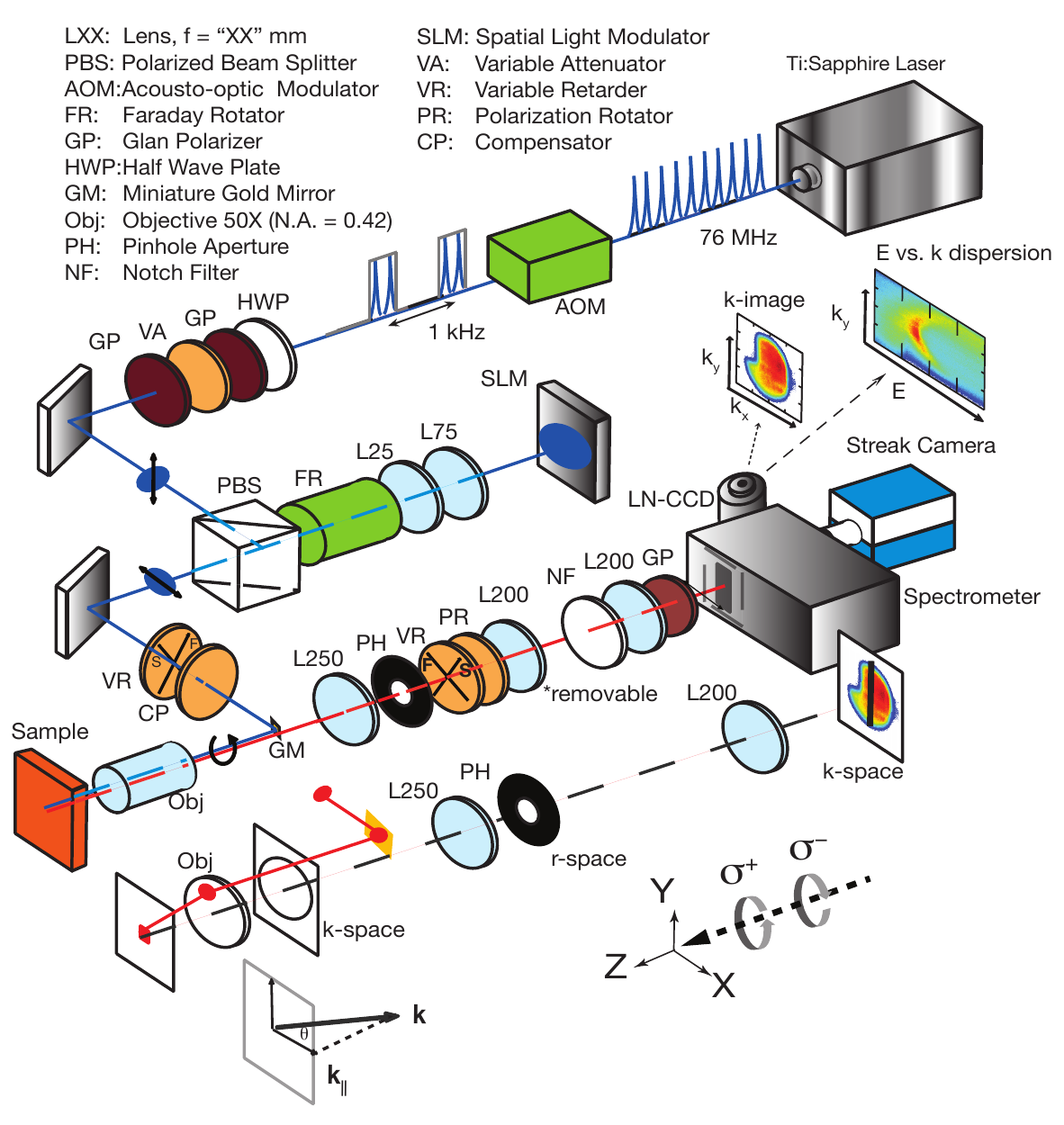}
\caption{\textbf{Experimental set-up: angle-resolved spectroscopy and polarimetry.} Luminescence from the microcavity is collected by a microscope objective (N.A. = 0.42, effective focal length $f_0 = 4$ mm). The collection angle is up to $\theta \approx 25^\circ$ in air. The luminescence with field distribution $F(X,Y)$ is Fourier transformed into a far-field image in the back focal plane of the objective with coordinates $(u,v)=(f_{o}\times \sin(\theta_{X}), f_{o}\times \sin(\theta_{Y}))$. This plane is mapped into k-space with $(k_X, k_Y) = k \times (u/f_o, v/f_o)$, where $k$ is the wavenumber and $k_X$ or $k_Y \equiv k_\parallel$, the in-plane momentum. The field distribution in this plane is relayed to the entrance plane of the spectrometer using a pair of lenses L250 ($f = 250$ mm) and L200 ($f = 200$ mm). In the conjugate real-space imaging plane (r-space), we place a 600-$\mu$m diameter circular aperture to spatially isolate luminescence and a single transverse mode within a circular $\approx$10-$\mu$m diameter area on the sample. The image at the entrance plane can be directed to the LN-CCD for time-integrated imaging/spectroscopy or to the streak camera system for time-resolved measurements. In this configuration, we measure the angular distribution of luminescence as k-space images (k-images) or spectra ($E$ vs. $k_\parallel$ dispersions) using respectively the 0-order or 1st-order diffracted light from the grating. By inserting a removable 200 mm focal-length lens (L200), we project the r-space luminescence to the entrance plane of the spectrometer.}
\label{fig:exp_setup}
\end{figure}

\begin{figure}
\includegraphics[width= 1.0 \columnwidth]{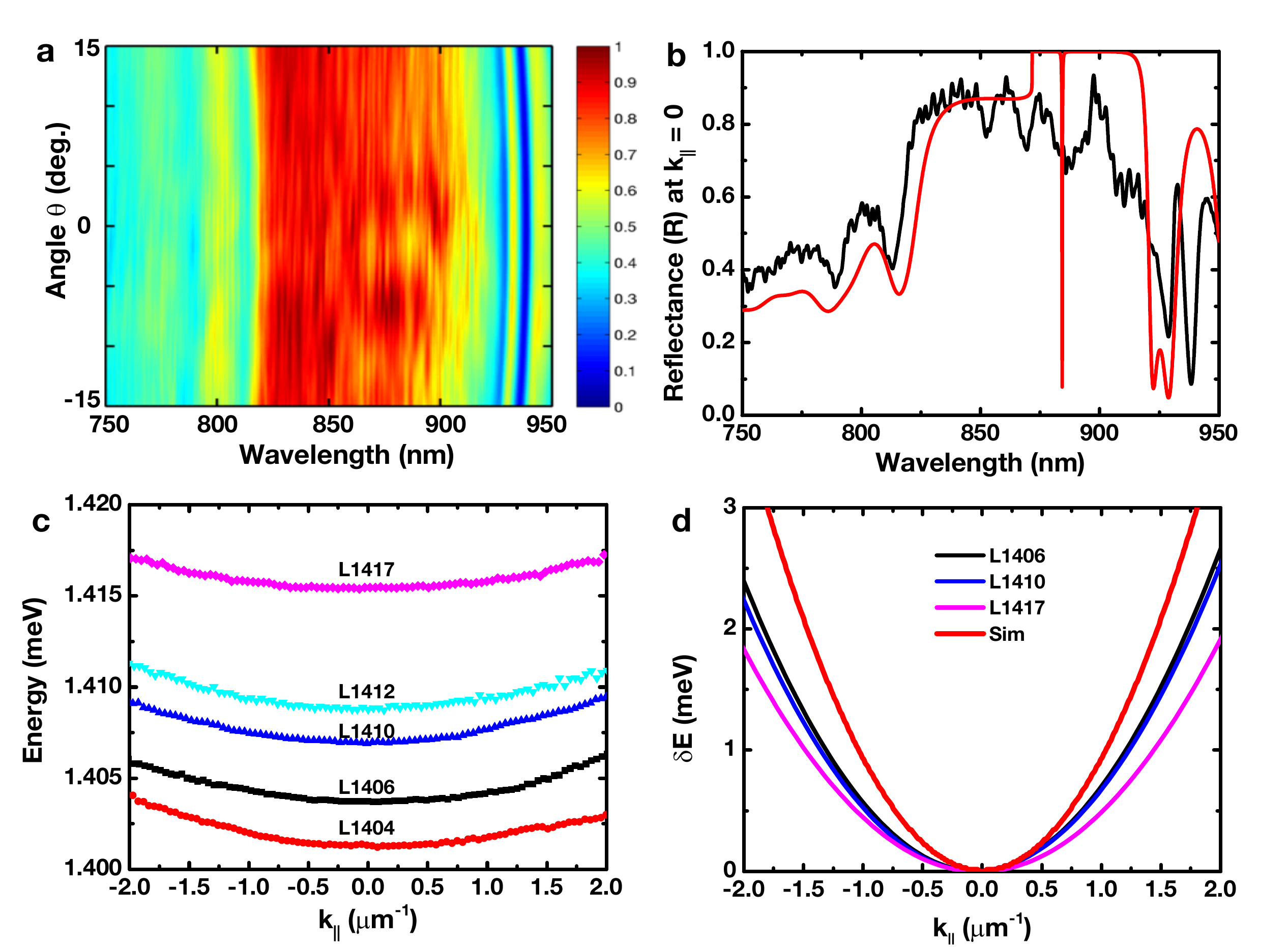}
\caption{ \textbf{Reflectance and dispersion.} \textbf{a}, Angle-resolved reflectance spectra from the front surface of the microcavity Sample-L1406. The false color represents the measured reflectance evaluated relative to that of a gold-coated mirror. \textbf{b}, Reflectance spectrum at $k_\parallel = 0$: measured (black solid line) and simulated (red solid line). \textbf{c}, $E$ vs. $k_\parallel$ dispersion curves determined from the angle-resolved luminescence spectra using peaks near the InGaAs-MQW bandgap at pump $P = 0.8 \, P_{th}$. \textbf{d}, $\delta E$ vs. $k_\parallel$ dispersion curves (parabolic fittings of the data shown in \textbf{c}), where $\delta E=E-E(k_\parallel=0)$. The red solid line is the simulated dispersion curve. The simulation is performed via a transfer matrix method using the structure shown in Fig. S1, including the optical absorption in GaAs layers but excluding exciton and free-carrier polarization.}
\label{fig:rt_ek}
\end{figure}

\begin{figure}
\includegraphics[width= 1.0 \columnwidth]{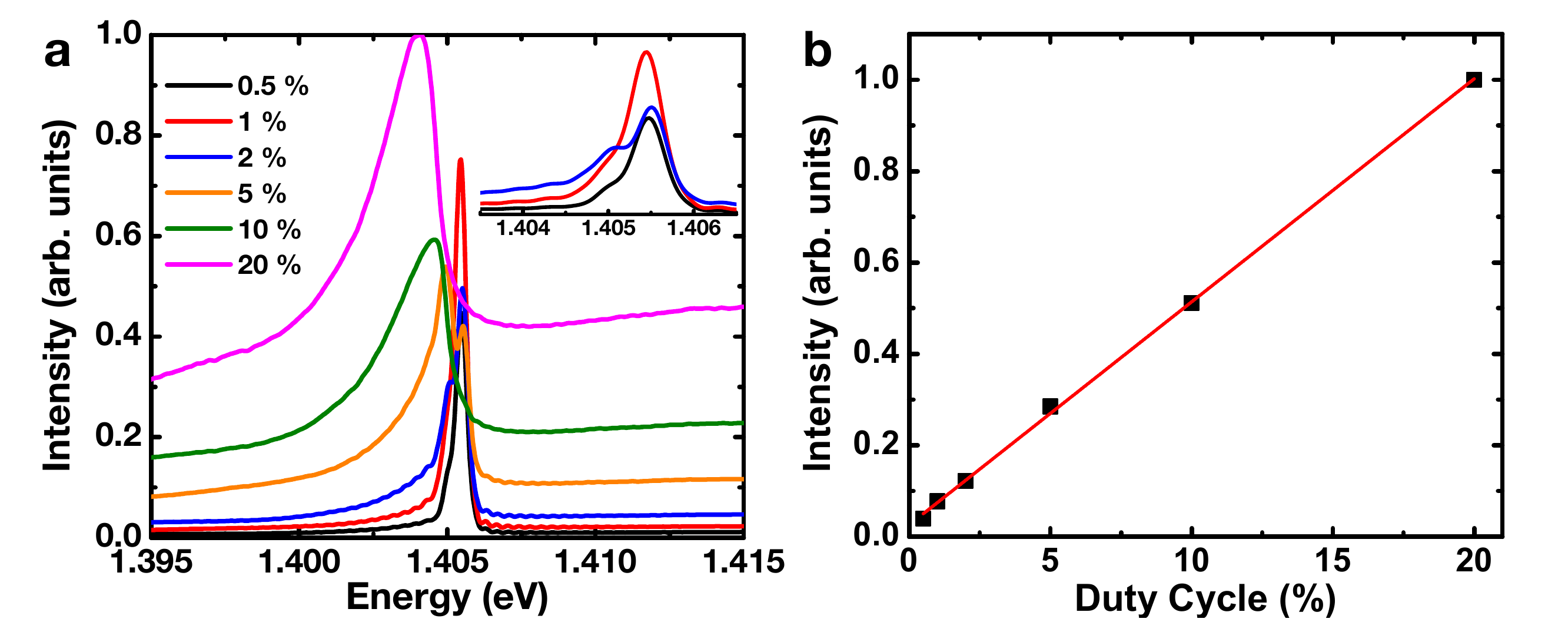}
\caption{ \textbf{Thermal management: temporal intensity modulation.} \textbf{a}, Time-integrated spectra at $k_\parallel = 0$ and at $P = P_{th}$ for an intensity modulation duty cycle from 0.5 to 20 $\%$. The inset shows the spectra for a duty cycle from $0.5\%$ to $2\%$. Thermal carrier heating has a detrimental effect on the laser action. The spectrally narrow radiation redshifts and broadens significantly for a duty cycle  $>$$2\%$. A spectrally narrow (linewidth $\Delta E<$ 0.3 meV) laser mode is only observed for a duty cycle below $1\%$. \textbf{b}, Spectrally integrated flux vs. duty cycle (black dots) and a linear fit (red line). Overall, with an increasing duty cycle, the luminescence redshifts and broadens spectrally with an almost constant radiative recombination efficiency. The directional radiation also becomes angularly broad (not shown), indicative of the reduction of spatial coherence. The time-averaged power transmitted into the microcavity at $1\%$ duty cycle and at $P_{th}$ is $\approx$0.1 mW.}
\label{fig:modulation}
\end{figure}

\begin{figure}
\centering
\includegraphics[width= 0.6 \columnwidth]{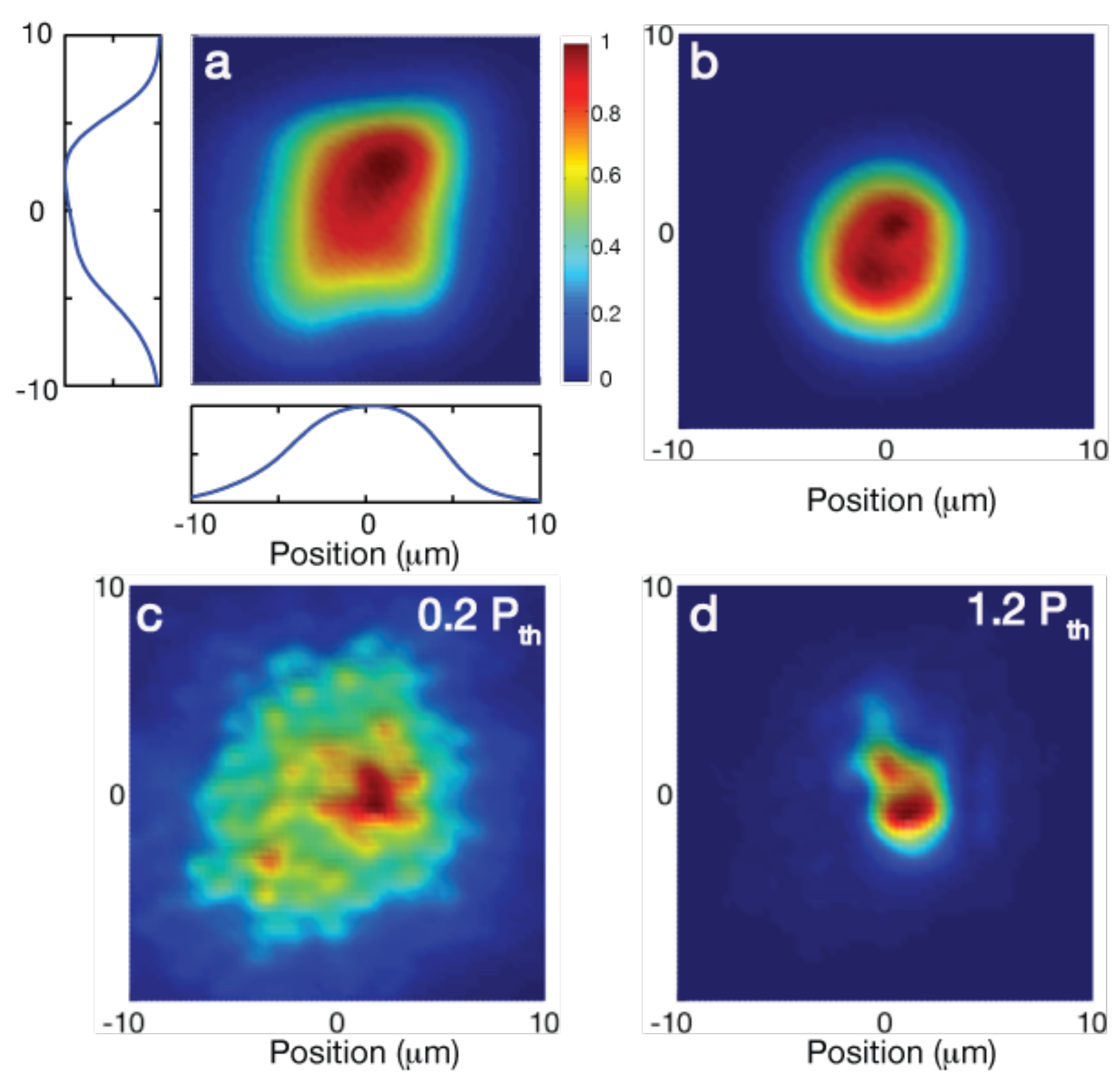}
\caption{ \textbf{Excitation beam profile and r-space luminescence image.} \textbf{a}, A real-space (r-space) luminescence image and cross-sectional profiles of a flat-top beam at the focal plane of the objective. The luminescence image is taken from the surface of a bulk GaAs sample and represents the beam profile. The $1/e^2$ width of x- or y-central cross section is about 18 $\mu$m, corresponding to about 250 $\mu m^2$ area. \textbf{b} Same as (a), but through a 10-$\mu$m diameter aperture. \textbf{c}, An r-space luminescence image from the microcavity surface below the threshold ($P = 0.2 \, P_{th}$). The luminescence intensity distribution is spatially inhomogeneous owing to disorder. \textbf{d}, An r-space image of a single transverse mode in the microcavity above threshold ($P = 1.2 \, P_{th}$). The Ti:Sapphire pump laser beam with a Gaussian profile is reshaped into a square flat-top beam profile at the focal plane of the objective by a phase-only two-dimensional (2D) liquid crystal spatial light modulator (SLM) (Holoeye PLUTO-NIR, 1920$\times$1080 pixels, pixel pitch = $8.0$ $\mu$m).  We isolate a single transverse mode by taking k-space images and spectra through a 10-$\mu$m diameter aperture positioned at the conjugate imaging plane, as shown in \textbf{d}. All k-space images and spectra reported in this article are taken through the aperture.}
\label{fig:beamprofile}
\end{figure}

\begin{figure}
\includegraphics[width= 1.0 \columnwidth]{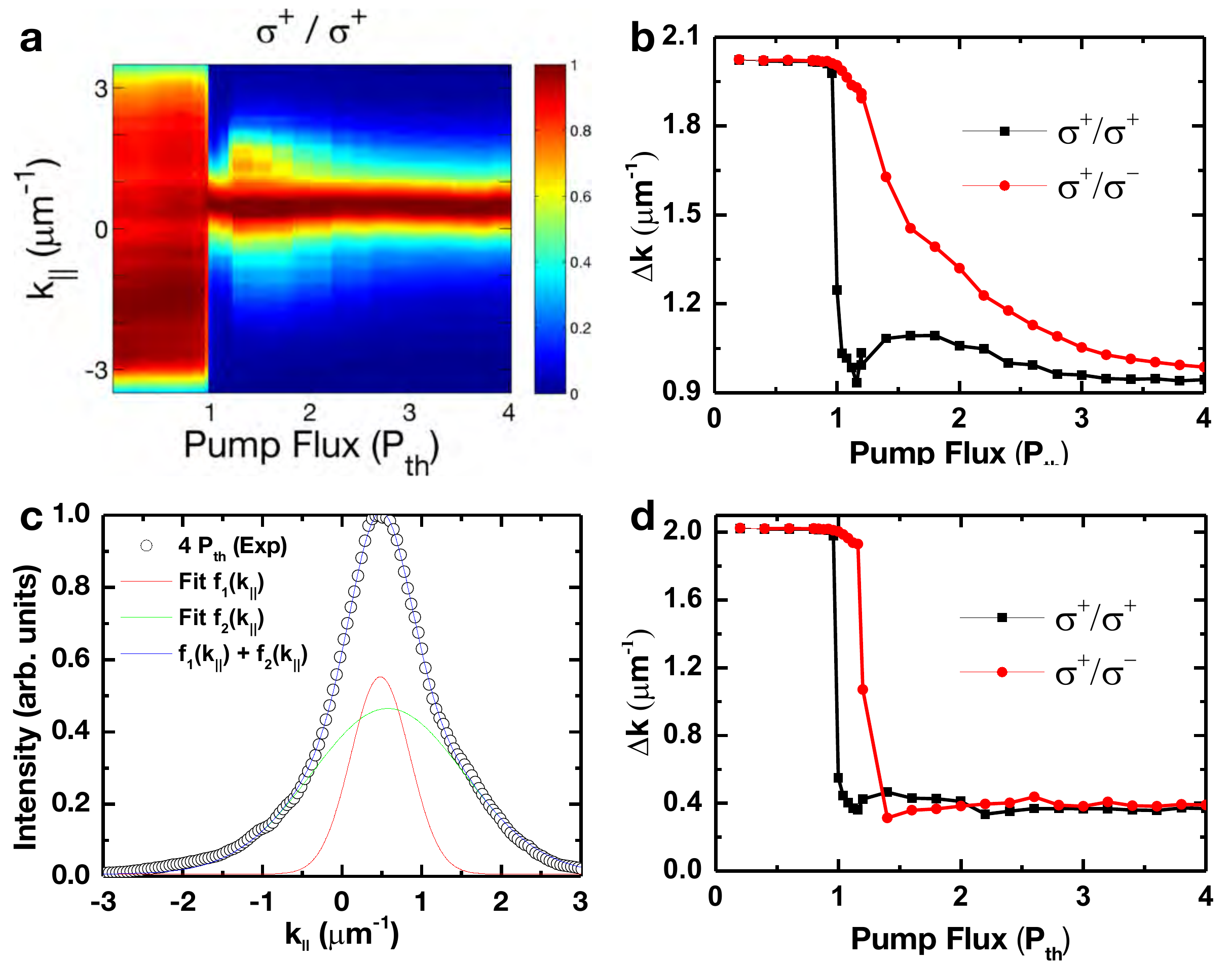}
\caption{ \textbf{k-space distribution and spatial coherence} \textbf{a}, Normalized cross-sectional profiles (along $k_Y-$axis at $k_X = 0$) of the k-space images (co-circular $\sigma^{+}/\sigma^{+}$ components) as a function of pump flux. The observable in-plane momentum is limited to a range from $\approx$ -3 to 3 $\mu$m$^{-1}$ by the objective (N.A. = 0.42). \textbf{b}, The standard deviation $\Delta k$ calculated from cross-sectional profiles of the co-circular (black solid squares) or cross-circular (red solid circles) component. The angular spread ($\Delta k$) decreases only by a factor of two at threshold. Such a large angular spread above threshold is due to the presence of an incoherent angularly broad background. \textbf{c}, A cross-sectional profile at $P = 4 \, P_{th}$ (black open circles) with double Gaussian fit results (red, green, and blue lines). A directional mode (central lobe, red line) emerges from the angularly broad background (green line). \textbf{d}, $\Delta k$ of the central lobe determined by double Gaussian fits. $\Delta k_\parallel$ decreases by a factor of 5 from 2 to 0.3 $\mu$m$^{-1}$ at threshold and remains constant with increasing pump flux. $\Delta k_\parallel = $0.3 $\mu$m$^{-1}$ corresponds to a coherence length $r_c \approx$ 4 $\mu$m.}
\label{fig:K_moment}
\end{figure}

\begin{figure}
\includegraphics[width= 1.0 \columnwidth]{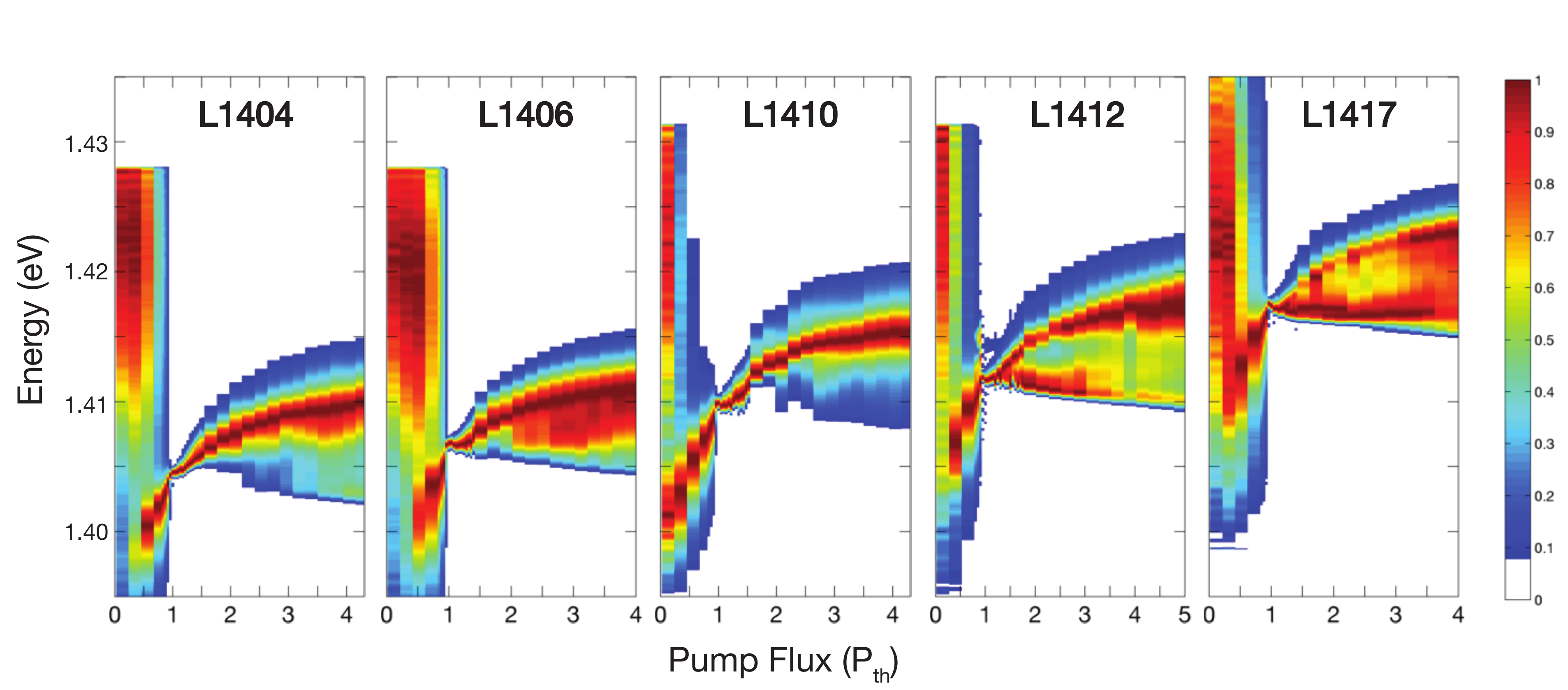}
\caption{ \textbf{Comparison of samples: luminescence spectra at $k_\parallel = 0$.} Normalized spectra of the co-circular component $\sigma^{+}/\sigma^{+}$ at $k_\parallel = 0$ vs. pump flux for five samples with detuning $\Delta =  E_X - E_C$ ranging from about $-5$ to $+5$ meV. The false color is in a linear scale. All samples lase at a threshold pump flux that varies by up to 50$\%$ and are labeled according  to the lasing energy (in meV) at threshold. The cavity resonance is estimated to be at 1.410 eV. A detailed analysis of these lasing characteristics will be reported elsewhere.}
\label{fig:2dsr}
\end{figure} 

\begin{figure}
\includegraphics[width= 1.0 \columnwidth]{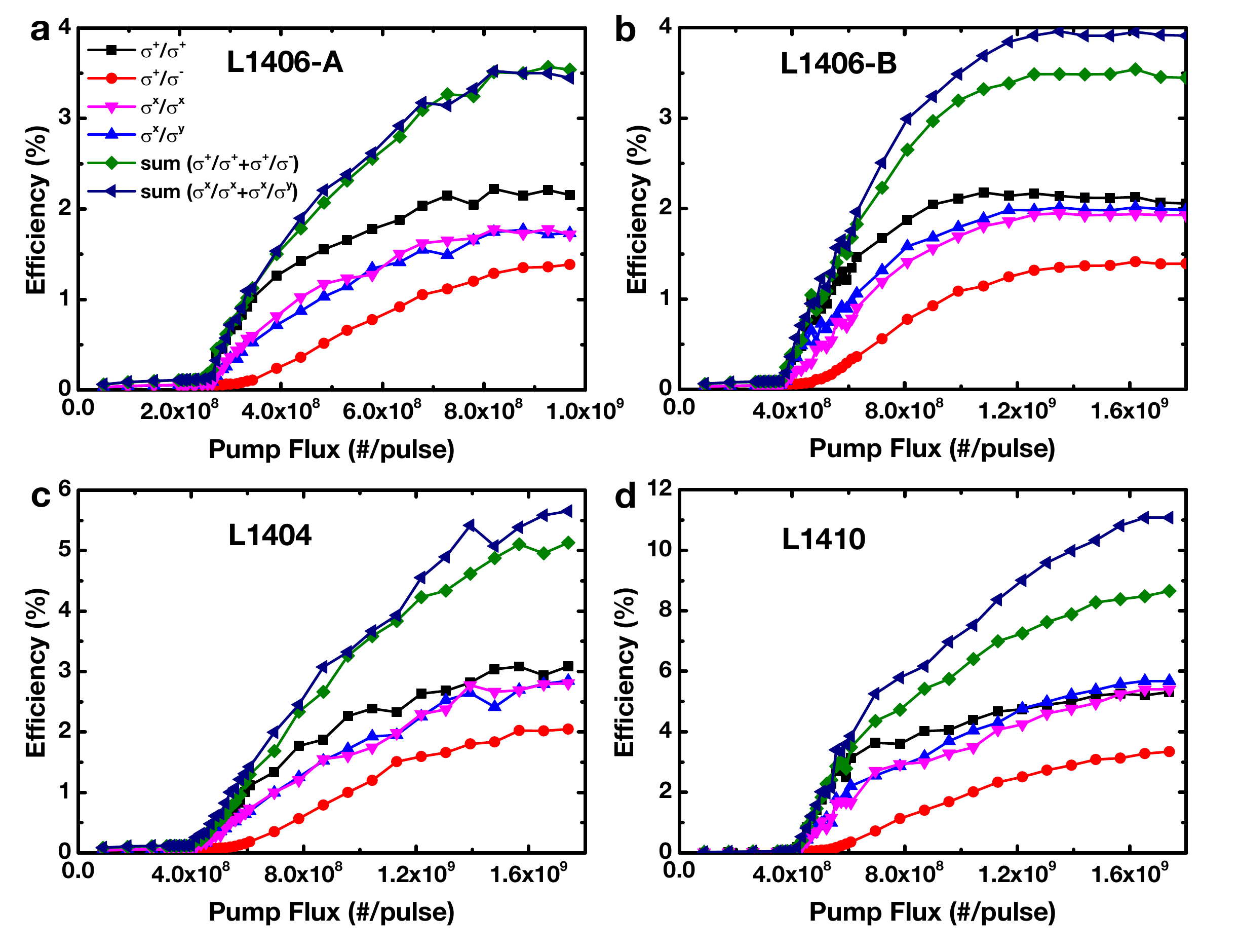}
\caption{ \textbf{Emission efficiency: Sample-L1404, Sample-L1406, and Sample-L1410.} Radiative emission efficiency vs. pump flux under a circulary ($\sigma^+$) or linearly ($\sigma^X$) polarized pump: \textbf{a}, Sample-L1406 (L1406-A mode); \textbf{b}, same sample as in \textbf{a} but fo a different mode (L1406-B mode);\textbf{c}, Sample-L1404; \textbf{d}, Sample-L1410. In Sample-L1410, the bandgap of the InGaAs MQWs is tuned to be close to the cavity resonance, leading to an increase in the lasing efficiency to 8$\%$ ($\sigma^+$ pump) or 11$\%$ ($\sigma^X$ pump). For the mode L1406-A shown in \textbf{a} (also shown in Fig. 2), the intensities of the co- and cross-linear polarized components under a linearly polarized pump are between those of the co- and cross-circularly polarized components. In general, the lasing efficiency converges to the same plateu at $P\gtrsim4\,P_{th}$ for $\sigma^+/\sigma^-$, $\sigma^X/\sigma^X$, and $\sigma^X/\sigma^Y$ components.}
\end{figure}

\begin{figure}
\centering
\includegraphics[width= 0.8 \columnwidth]{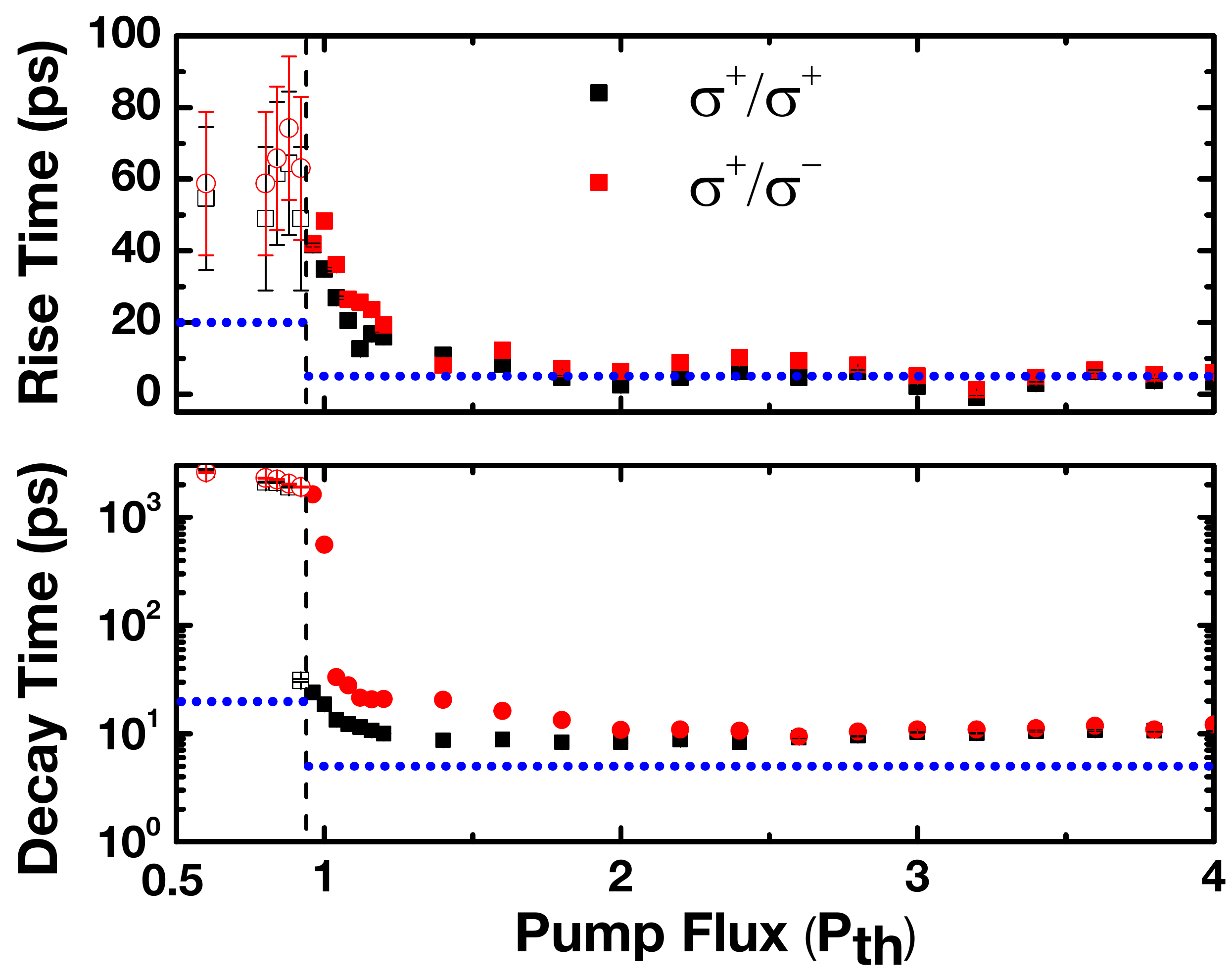}
\caption{ \textbf{Laser dynamics: rise and decay times.} The rise and decay times of co-circular ($\sigma^{+}/\sigma^{+}$, solid black squares) and cross-circular ($\sigma^{+}/\sigma^{-}$, solid red circles) components above threshold. Note that the y-axis of the decay time sub-figure is in a log scale. The time zero is determined from 2-ps pump pulses specularly reflected off the microcavity sample and passed along an identical optical path as the luminescence. The short-dot lines represent the 5-ps or 20-ps temporal resolution. The 20-ps temporal resolution below threshold is due to a decreased sweep speed used to capture luminescence up to about 3 ns below the pump flux indicated by the black dash line. The rise time is determined at the intensity peak of the time traces. The decay time is determined by an exponential decay fitting. The rise time decreases from about 70 ps to below 10 ps for $P > 1.5 \, P_{th}$. The decay time decreases from above 2 ns to below 10 ps across the threshold.}
\label{fig:rsie_life}
\end{figure}

\begin{figure}
\includegraphics[width= 1.0 \columnwidth]{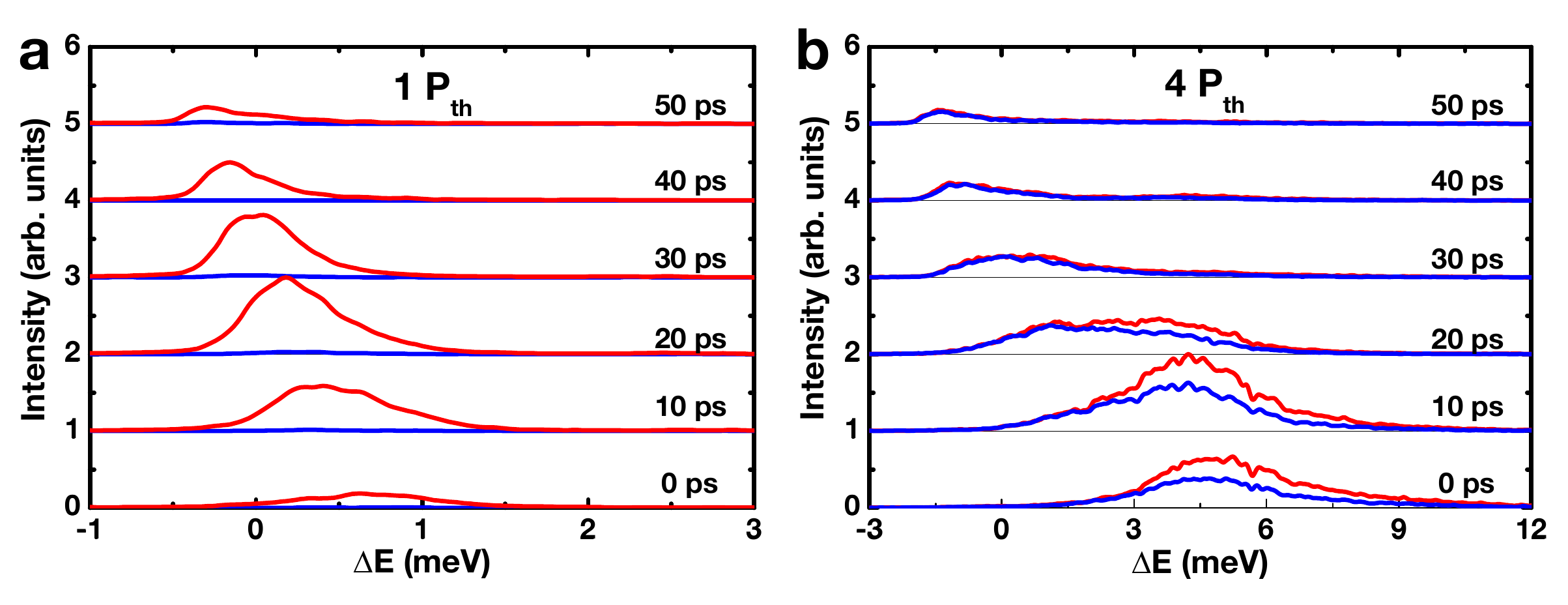}
\caption{ \textbf{Transient lasing spectra.} Cross-sectional transient spectra at specified time delays extracted from the temporally and spectrally-resolved streak images in Fig. 3b:  \textbf{a}, $P = P_{th}$; \textbf{b}, $P = 4 \, P_{th}$. Transient spectra are averaged over 5 ps and normalized to the maximal peak intensity of the co-circular component. The spectra are equally scaled but offset vertically by 1. Red lines represent the co-circular ($\sigma^{+}/\sigma^{+}$) component, while blue lines represent the cross-circular ($\sigma^{+}/\sigma^{-}$) one. The energy scale is measured with respect to to 1.406 eV ($\lambda = 882 \, nm$), which is the peak lasing energy at the threshold $P_{th}$. Note the different scales of the x-axes in \textbf{a}  and \textbf{b}. At $P = P_{th}$, the co-circular component maximizes at a time delay of 20 ps. The emission remains highly circularly polarized after 50 ps. At $P = 4 \, P_{th}$, the emission is spectrally broad and blueshifts  by about 5 meV near zero time delay. A spin-dependent energy splitting of about 1 meV can be observed for delays less than 20 ps. The emission spectrum reaches a maximum at about 10 ps, and then gradually decreases in overall magnitude and redshifts with time. As the pump flux increases, a secondary radiation mode emerges at a lower energy after 20 ps. Such a low-energy mode exhibits a longer decay time ($\gtrsim30$ ps) and is less circularly polarized. We attribute this secondary mode to the recombination of electron-hole carriers that are photoexcited in the GaAs spacer layers but have diffused into the InGaAs MQWs.}
\label{fig:trsr}
\end{figure}

\end{document}